# Timekeeping infrastructure for the Catalina Sky Survey

Robert L. Seaman*[a], Alex R. Gibbs[a]

[a]Lunar and Planetary Laboratory, University of Arizona
1629 E. University Blvd., Tucson, AZ, USA 85721

## ABSTRACT

Time domain science forms an increasing fraction of astronomical programs at many facilities. Synoptic and targeted observing modes of transient, varying, and moving sources rely on precise clocks to provide the underlying time tags. Often precision is mistaken for accuracy, or the precise time signals never reach the instrumentation in the first place. We will discuss issues of deploying a stable high-precision GNSS clock on a remote mountaintop, and of conveying the resulting time signals to a computer in a way that permits hardware timestamping of the camera shutter (or equivalent) rather than the arbitrary delays encountered with non-real-time data acquisition software. Strengths and limitations of the Network Time Protocol will be reviewed. Timekeeping infrastructure deployed for the Catalina Sky Survey will serve as an example.

**Keywords:** Timekeeping, time, clocks, NTP, IRIG-B

## 1. INTRODUCTION

The Catalina Sky Survey[1] (CSS) is a NASA funded project to discover and carry out astrometric measurements of Near Earth Objects (NEOs), which are asteroids and comets that approach within 1.3 times the Earth's distance from the Sun. The Catalina project has been in operation for two decades[2] and has discovered about 46% of the 18,000+ known NEOs, including more than 600 close-approaching Potentially Hazardous Asteroids (PHAs) larger than 140 meters[3] and capable of extensive regional damage should they impact Earth, as well as more than 150 NEOs larger than a kilometer and posing a potential long-term global threat. Indeed, CSS is the only team to have discovered (thankfully small) asteroids before Earth impact, three times, in 2008, 2014, and 2018, over Sudan, the north Atlantic Ocean, and Botswana respectively.

Discovering a comet or asteroid involves a painstaking automated search through repeated multiple exposures of large patches of the sky many times the width of the full moon. Human eyes must validate thousands of candidate discoveries each night for moving points of light otherwise often indistinguishable from background stars. This results in two-dimensional positional measurements on the celestial sphere, in estimates of the objects' brightness in reflected sunlight, and in the dates and times corresponding to these observations. We will discuss the infrastructure that CSS has implemented for high-precision, reliable, and accurate timekeeping, necessary to the later calculation of orbits linking CSS observations to those from other professional and amateur asteroid observers.

Typically, an astronomical observation has been timestamped relative to the software used in data acquisition from Charge-Coupled Devices (CCDs) or other camera detector technology. Such software will ask the system time, then command the camera shutter to open, or an equivalent operation marking the beginning of an exposure for cameras lacking shutters. Sometimes multiple timestamps may be taken at different phases of the exposure process. However software works in each case, only very rarely are astronomical cameras controlled by real-time operating systems, rather they are more often read under Linux or Windows relying on standard, notoriously imprecise, computer clocks. A non-real-time OS cannot guarantee the absence of an arbitrary delay between timestamp and the beginning of an exposure.

The Catalina Sky Survey has instrumented the camera shutters on three of our telescopes to permit direct hardware time-capture measurements for the start and end of each image exposure. For those cameras using large scroll shutters we measure not just the start and end of each exposure, but also the start and end of the traverse motion across the CCD field of view when opening or closing the shutter. These hardware measurements are completely independent of timestamps resulting from the computer system clock, which are still recorded and can be compared during each exposure timeline. We further discuss the appropriate use of the Network Time Protocol[4] (NTP) to minimize random and systematic uncertainties in the computer system clock for those software timestamps that continue to be taken.

*seaman@lpl.arizona.edu; phone 1 520 621-4077; catalina.lpl.arizona.edu

## 2. TIME IN ASTROMETRIC OBSERVATIONS

Time-related issues are diverse and numerous in human society, especially in the sciences. Time in astronomy is used to point telescopes, to sequence observations, to tag time-series observations, and for large numbers of other purposes. In its reciprocal relationship as frequency, time appears in many other guises. In this document we will consider the particular case of astrometric observations, but almost all details will be applicable to a wide range of science use cases.

Figure 1 shows a typical NEO discovery observation from the Catalina Sky Survey. Descriptive metadata (the first nine rows in this case) are followed by 3 or 4 rows of measured celestial positions and apparent magnitudes at the corresponding times. A systematic or random error in the time may manifest as along-track positional uncertainties or vice versa.

```
COD G96
OBS R. A. Kowalski
TEL 1.5-m reflector + 10K CCD
NET GAIA-DR1
AC2 sixty@lpl.arizona.edu
MEA B. M. Africano, E. J. Christensen, D. C. Fuls, A. R. Gibbs, A. D. Grauer
MEA H. Groeller, J. A. Johnson, R. A. Kowalski, S. M. Larson, G. J. Leonard
MEA R. L. Seaman, F. C. Shelly
ACK 18Jun02.G96.S12109.1.75.mrpt
     ZLAF9B2   C2018 06 02.34330 16 11 10.54 -11 19 34.5          19.3 G      G96
     ZLAF9B2   C2018 06 02.34851 16 10 59.30 -11 20 21.6          18.2 G      G96
     ZLAF9B2   C2018 06 02.35895 16 10 36.59 -11 21 56.9                     G96
```

Figure 1. Discovery observations from the Catalina Sky Survey 60" telescope for the impacting asteroid 2018 LA. This is the 80-column MPC-1992 format for reporting astrometry to the Minor Planet Center. The observation time stamp appears as the decimal UTC calendar date in columns 16-32.

In 2015, Commission 20 of the International Astronomical Union adopted the Astrometry Data Exchange Standard (ADES)[5], which has continued to evolve through at least 2018. It is possible that the data format in Figure 1 will be supplanted by ADES 2018 by the time this document is released. This is of great interest to members of the international NEO community, but is largely an implementation detail for the remainder of our discussion. Whatever the standard format, data representation, or data typing issues involved in capturing observational timestamps, issues of reliability, precision, and accuracy remain the same and will benefit from being addressed using the techniques described here.

### 2.1 Precision clock system design

At a remote observatory site, implementing precision time signals is approximately as complex as implementing power or network infrastructure. Which is to say that utility power will itself be complicated by the need for generator backup, and networking by microwave links or expensive long-distance fiber runs that are subject to relatively frequent brief outages and occasional longer backhoe-related service interruptions. Similarly to mitigate outages and service degradation, reliable and accurate timekeeping will require deploying and managing a local reference clock, generally using a GPS or multi-constellation GNSS disciplined reference clock. Figure 2 shows the infrastructure CSS has implemented for its telescopes on Mount Lemmon and Mount Bigelow in the Santa Catalina Mountains in southern Arizona. The specific components are:

1. Clear-sky access to multi-constellation GNSS signals; CSS uses the U.S. GPS and Russian GLONASS satellites.

2. A GNSS antenna mounted on an external mast. A 20m coaxial cable connects that antenna to the GNSS receiver. CSS antennas have integrated lightning protection or this can be added in-line with the coax. An expensive option is to convert the signal to a proprietary fiber optic protocol. The spare receiver on the University of Arizona campus operates acceptably with an indoor antenna, but the geometrically odd metallic surfaces in telescope domes make indoor antennas not an option.

3. Modular GNSS clocks (receiver, CPU, OCXO clock, I/O modules, power supply). The CSS trade study selected Meinberg Lantime M1000[6] units, but there are similar quality clocks from other vendors. The reference clocks implement:

4. NTP served over the local subnet.

5. DCLS IRIG-B (IEEE 1344) time codes via BNC coax.

6. TTL to ST fiber optic converter[7]; matching FO to TTL media converter[8]
7. IRIG-B PCIe card[9] with TCXO clock, which implements:
8.    TTL time capture feature
9. via another TTL to FO to TTL converter pair from shutter controller box
10. To assign precise mid-exposure times to each pixel in the image focal plane.
11. IRIG-B card also generates 1PPS signal that can be monitored or be used as its own reference clock.

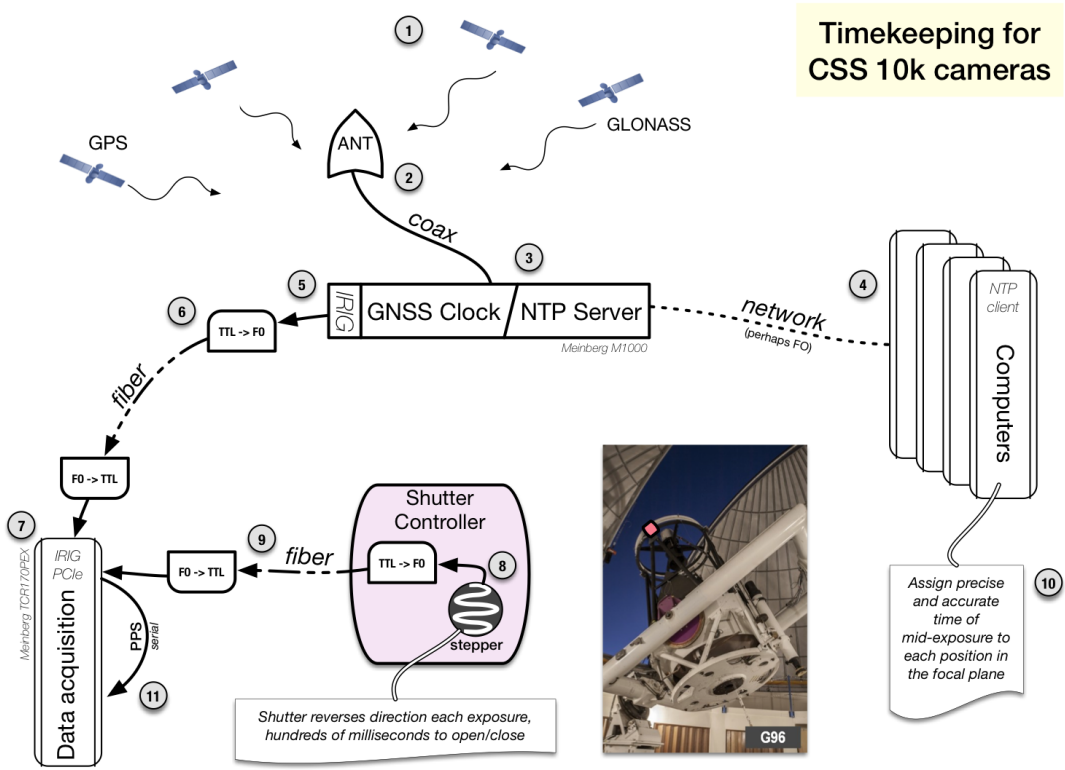

Figure 2. Precision timekeeping infrastructure deployed at the Catalina Sky Survey telescopes on Mount Lemmon and Mount Bigelow, Arizona. A multi-constellation GNSS-disciplined clock provides a local NTP stratum-1 reference as well as serving IRIG-B DCLS time codes via fiber optic links. Hardware time capture is implemented by PCIe IRIG cards installed in our CCD data acquisition computers. Numbered elements are described in the text.

**2.2 Hardware implementation**

Catalina's GNSS clocks on Mount Lemmon and Mount Bigelow are deployed in the 6U rackmount enclosures as shown in Figure 3. The unit is divided into roughly four parts. First is the modular reference clock itself, a Meinberg M1000 with a multi-constellation GNSS reference clock that supports both the U.S. GPS and Russian GLONASS satellites. The current version of the receiver also supports the European Galileo satellites as well as the Chinese Beidou. Other major vendors such as Spectracom and Microsemi offer similar modular designs with various trade-offs. One advantage of Meinberg is unencumbered access to software updates; a software licensing contract is more typical.

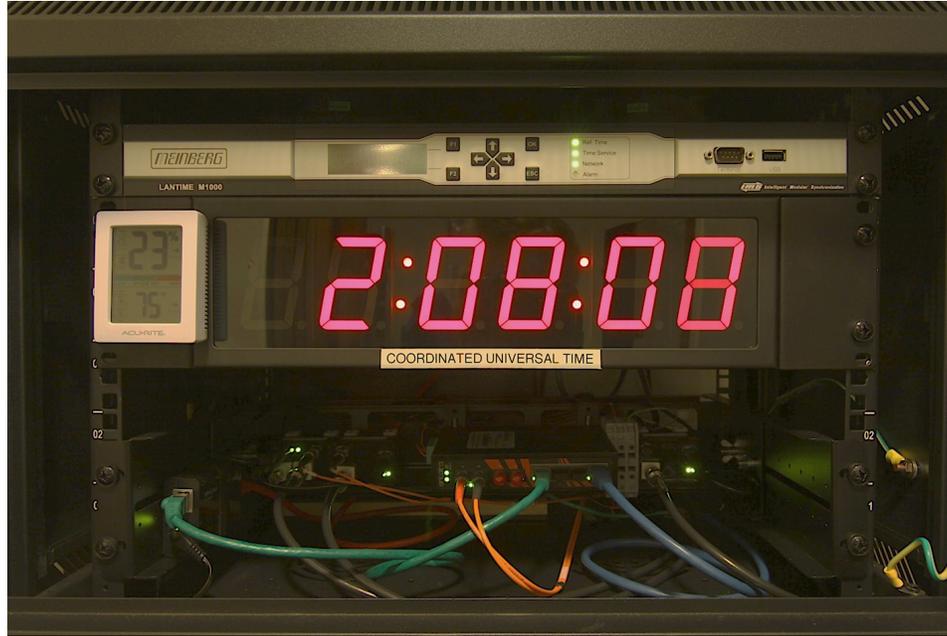

Figure 3. Grounded enclosure for GNSS reference clock (top) and related hardware, to allow timekeeping infrastructure to remain powered during lightning shutdown at observatory. All network and IRIG-B signals are fiber-isolated (bottom). Small UPS powers the enclosure. GNSS antenna has built-in lightning protection.

The second part is the enclosure itself which has metal panels on five sides and a glass door on the front. Panels on all sides allow easy access from any direction and the addition of a grounding kit bonds all panels together to provide lightning protection in the mountaintop environment. We have grounded the units through 10-gauge wires into the building ground; consult your local electrician for appropriate wiring. The enclosure also includes a small dedicated UPS sized for about a one hour outage. A pair of variable-speed fans cool the enclosure through the top of the box with an inexpensive digital thermometer to monitor the environment, including ambient humidity. The enclosures are wall-mountable or the addition of feet or a couple of short lengths of 2"x3" lumber provide for both ventilation and the routing of cables below.

Third is an extensive set of fiber converters for the various network, IRIG-B, and 1PPS signals that exit and enter the box. Aside from a single power cord for the UPS, the equipment ground, and coax for the GNSS frequency radio antenna, only fiber optic connections are made to the box, both for lightning protection as well as easy long distance connections between domes on the different mountain tops. In particular the TTL to FO to TTL converters are rated for about 2 km and it is convenient to use visual confirmation of the 100 Hz IRIG-B signals leaving the box or the 1PPS (one pulse per second) heartbeat returning from the PCIe devices.

And fourth and final is the addition of a power-over-ethernet (POE) SNTP-driven digital UTC clock[10]. This provides immediate confirmation that the reference clock continues to work since without a reference the display flashes, and also that the box remains powered since everything but the POE clock is on uninterruptable power. If the power to the box goes out the UTC display will go dark but the various LED health lights will remain lit. We have chosen to deploy red UTC clocks, but also similar white local time displays elsewhere in the telescope control rooms. These units are programmable via a simple web interface. The FO network media converter supports a few switched copper ports; our units do not directly support POE, so an inexpensive POE injector is also necessary.

**2.3 Electromechanical shutter time signals**

A requirement for timing any physical phenomenon is to arrange for a contemporaneous pulse suitable for time capture. Generally this means an electronic pulse, and for the Meinberg time capture equipment CSS uses, the requirement is for a pulse with a falling edge. Figure 4 is a four-channel oscilloscope trace connected to the various signals from our 10K CCD scroll shutter simulator. Takeaway #1 for this paper is that time is something to be measured, not estimated.

These camera shutters were manufactured in-house. The opening in the shutter moves from left-to-right to open and then continues from left-to-right to close at the end of the exposure. For the next exposure the shutter will reverse direction, scrolling from right-to-left to open and then right-to-left to close again. Subsequent pairs of exposures repeat the pattern.

The advantage of such shutters is a precisely equal exposure duration across the detector. Some care is taken to accelerate to a steady motion before the shutter clears the first row or column of pixels. Knowledge of the axis of motion and of the direction of travel will allow correcting the time at every pixel position on the focal plane.

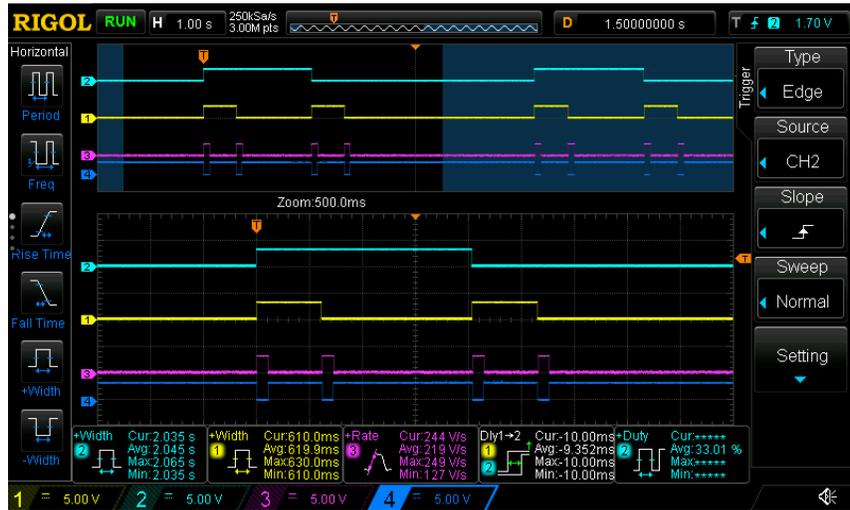

Figure 4. Oscilloscope trace of CSS camera shutter simulator. Green signal is per-exposure shutter signal from typical CCD controller. Yellow is the motion flag from the stepper motors driving the Catalina 10K scroll shutters, about 0.6s per traverse. Red is a pulse coincident with motion start and end. Blue is the inverted pulse required by falling-edge time capture.

The green trace in figure 4 represents the camera's exposure signal, generally created by the CCD head electronics. A signal is raised when the detector begins integrating, and falls during readout and the subsequent CCD prepare cycle, telescope slew, etc., in particular including a guard time for the overhead of the shutter traversal.

The yellow trace is generated by the CSS shutter controller box, and in particular by the microcode program created in-house for a commercial stepper motor controller. This signal is high when the stepper motor is actually moving the shutter in either direction and either opening or closing. A Schmitt trigger makes sure both the motion ramp up and ramp down create a sharp edge since these were significantly asymmetrical and otherwise fluctuate in a phenomenon similar to ringing or debouncing a switch. A single open or close motion takes 0.6 seconds to complete.

The red trace is a pulse created for both edges of the yellow trace. And the blue trace is simply the red trace inverted to create the falling edge required by the Meinberg devices. Commercial ICs can perform a variety of such signal conditioning operations as might be needed for your application.

## 2.4 Computing timing metadata

Time capture measurements are only of use after they have been recorded in some fashion and associated with the particular camera exposure or other experimental or instrumental activity.

```
Keyword          Calculated                   Event              Time capture

MJDOPBEG=   57539.254673156218 / [d] Open shutter:  2016-05-31 06:06:43.7606972
MJDOPEND=   57539.254680077422 / [d] Open ended:    2016-05-31 06:06:44.3586895
MJDMID  =   57539.254908091840 / [d] Mid exposure:  2016-05-31 06:07:04.0591350
MJDCLBEG=   57539.255136106265 / [d] Close shutter: 2016-05-31 06:07:23.7595814
MJDCLEND=   57539.255143026254 / [d] Close ended:   2016-05-31 06:07:24.3574683
EXP_MEAS=           39.9988841 / [s] Measured exposure duration
OSHUTDUR=            0.5979920 / [s] Measured shutter open duration
CSHUTDUR=            0.5978870 / [s] Measured shutter close duration
```

Mid-exposure for each pixel Y location ← Shutter Direction (reverses) ← Physical shutter speed (+ converging optics) ← Mid-exposure for the middle pixel of the detector

Figure 5. FITS header keywords measured or calculated by captured shutter events.

Figure 5 shows the FITS[11] keywords created as a result of the per-exposure CSS 10K camera shutter time capture events. The Meinberg software API returns the results of several time capture events when called after the exposure. The value of these timestamps can be seen as the FITS comment at the end of the *BEG and *END keyword, e.g., "2016-05-31 06:06:43.7606972". Before each exposure a call to the same API drains the time capture ring buffer. A script with knowledge of the details of our shutters handles various edge cases such as non-scroll shutters that produce only a single open and single close time capture.

The tenth microsecond precision of the time capture is achievable when performed by the reference clocks themselves (and indeed, 10 MHz signal handling is widely supported throughout the commercial precision timekeeping industry). The TCR170pex PCIe cards that CSS uses in its data acquisition computers are themselves rated to 5 microseconds and this should be regarded as the achievable precision in this configuration. The mechanical repeatability of the CSS scroll shutters is about 1 millisecond. A typical Uniblitz iris shutter has about a 10 millisecond asymmetry between opening and closing, against and with the spring loading, respectively. The final precision retained for FITS keywords and database columns is some combination of all of these factors.

The ISO-8601 time capture strings are converted to Modified Julian Date format to simplify arithmetic and the shutter open and close duration are measured (symmetrical to about 0.1 milliseconds for the scroll shutters) and the actual shutter duration is computed (39.999 seconds compared to the requested 40 seconds). Finally the MJD of mid-exposure for the middle of the detector is computed.

As shown in red, knowledge of the physical shutter speed and direction can be used to calculate the mid-exposure timestamp for every pixel in the focal plane of the camera.

Other instrumentation may require timing different events, for example a shutter-less CMOS camera might nevertheless expose one or more timing signals that can be converted to precise timestamps of the camera workflow through simple arithmetic and logistics. The Meinberg devices support dual time capture channels and other vendors multiple channels. Only a single channel is needed if the timing events are constrained to happen sequentially at least up to the limit of the event ring buffer, but complex instrumentation or experimental setup may interleave multiple classes of events.

**2.5 Shared timing infrastructure**

Once timing infrastructure as discussed in sections 2.1 and 2.2 is deployed at an astronomical site, it is straightforward to share the precision time signals, e.g., NTP or IRIG-B, to additional hardware or software systems.

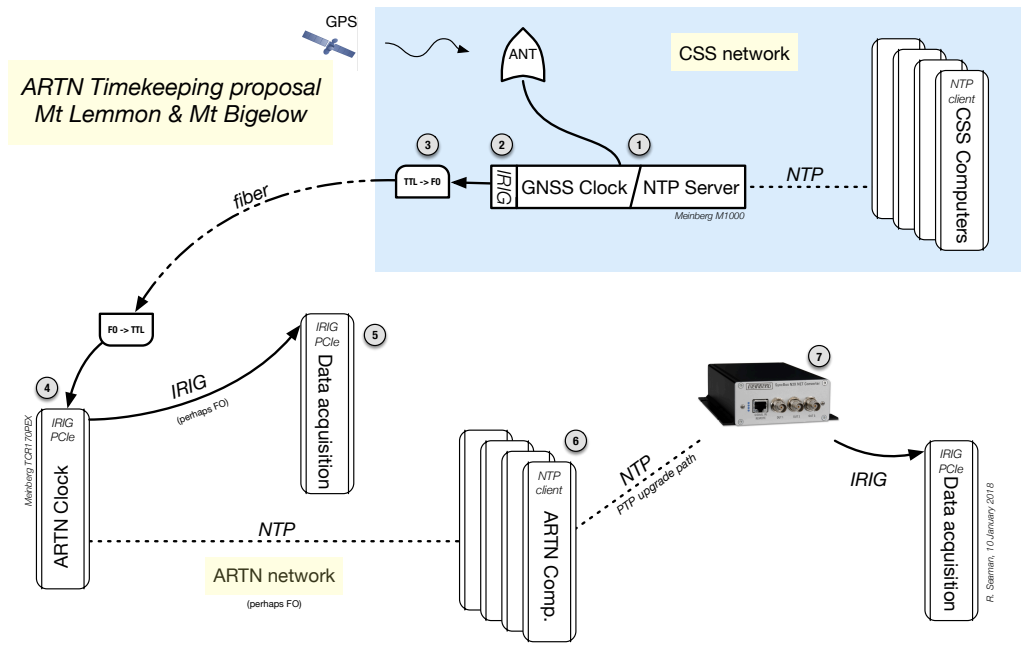

Figure 6. It is straightforward to extend NTP or IRIG-B time signals to additional telescope domes or associated computers. An IRIG-B PCIe card in a computer isolated on a separate subnetwork can serve as an isolated NTP stratum-0 reference clock.

The Catalina Sky Survey telescopes are located at astronomical sites operated by Steward Observatory of the University of Arizona and serving many other telescopes. Several of these telescopes are part of the Arizona Robotic Telescope Network project that unites their operation layered on top of a core set of technologies such as telescope control and automation. Timekeeping fits well in such a paradigm.

As shown in figure 6, a single fiber supplying IRIG-B DCLS time codes (http://irig.org, as implemented by each vendor) is sufficient to convey precise time from one telescope to neighboring domes. By connecting the fiber to a device such as the Meinberg TCR170pex PCIe card, the clocks are synchronized and perhaps daisy chained one dome to the next. The new computer's local reference can then serve as an NTP stratum-0 refclock, perhaps on an entirely different sub-network. Devices such as Meinberg's SyncBox/N2X[12] can allow converting downstream NTP back into IRIG-B signals and time capture wherever the new network reaches.

This configuration supports a variety of upgrade paths, including layering on the Precision Time Protocol[13] (PTP) that is optionally supported by various of these devices instead of, or in addition to, IRIG-B.

## 3. TIMEKEEPING AS AN APPLIANCE

There are many clocks resident at a typical astronomical site and the overriding goal is to ensure that all clocks are synchronized at the level(s) required for various tasks. This first goal is equivalent to measuring time as with any other scientific quantity. The second goal then is to monitor the quality of these measurements, and this is also rendered straightforward by timekeeping vendors.

### 3.1 Monitoring multiple clocks

Once time signals such as NTP, PTP, or IRIG-B have been distributed among diverse observatory subsystems, it is relatively straightforward to monitor these several clocks and generate alerts should the clocks exceed preset limits.

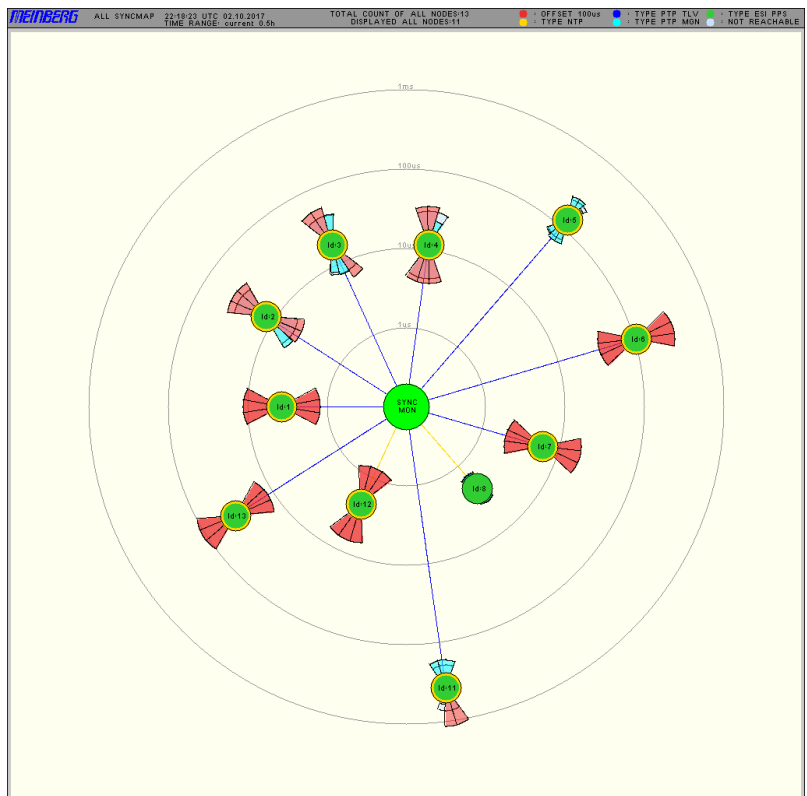

Figure 7. The CSS Meinberg GNSS reference clocks include a feature for easily monitoring an ensemble of NTP servers as well as IRIG-B disciplined references (via their 1PPS outputs). In this display, the concentric circles are a log-polar representation of each clock's offset from the GNSS reference. The outer circle is one millisecond and the innermost is a one microsecond offset.

Figure 7 shows the Meinberg Syncmap interface as configured for one of the Catalina mountaintop installations. Each symbol represents a different computer or timekeeping devices on a polar log plot ranging from a 1 millisecond offset at the outer circle to 1 microsecond at the innermost circle. The red and blue flags attached to each point convey information about the variation of each clock and whether fast or slow. All points depicted are NTP instances except for the device with very small variation at about 4 microseconds offset to the lower right which is the 1PPS signal from the IRIG-B clock embedded in the CSS 10K camera's data acquisition computer. A Meinberg modular I/O card is required to access these offsets.

Alerts can be set for each host such that an NTP source might alarm if it exceeds 1 millisecond offset and a 1PPS source if it exceeds 10 microseconds, for instance. A variety of other alert types are supported. Clicking on each symbol will create plots such as figure 8-13 below.

### 3.2 Embedded reference clocks

Some systems are more mission critical than others. In the case of minor planet astrometry, the most important timing information is related to timestamps attached to data acquisition. Figure 8 shows a two-week period of monitoring the stratum-1 NTP server layered on the IRIG-B driven stratum-0 reference clock in the data acquisition computer of Catalina's Schmidt telescope (MPC code 703). This uses the standard shared memory interface built into NTP and a simple NTP configuration. The red trace here represents the Linux system clock achievable using these straightforward techniques on a relatively good Dell desktop class machine. The green trace is ground truth supplied by the GNSS reference clock.

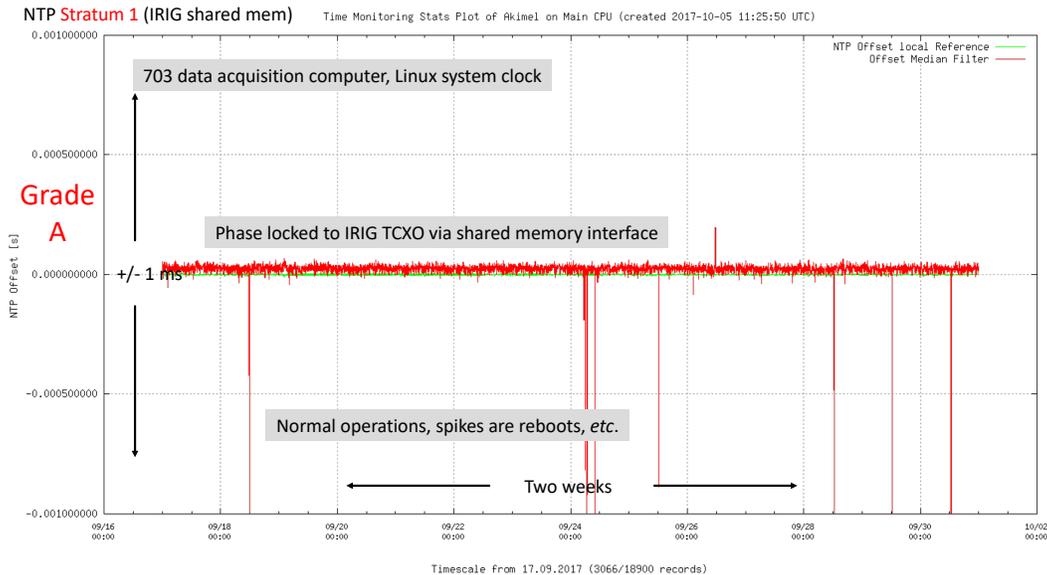

Figure 8. Example of an NTP stratum-1 server referencing an IRIG-B PCIe card as a shared memory clock. The x-axis is a typical two week period and the y-axis is +/- 1 ms. Spikes are instances of daytime network resets, and other well-understood interruptions to normal operations and are otherwise negligible.

### 3.3 Local NTP reference clocks

The Network Time Protocol is a useful tool for mid-precision technical and science use cases. The challenge is to ensure an accurate representation of the appropriate time standard to whatever precision is claimed. Figure 9 shows a similarly good quality Linux desktop host configured as a typical NTP stratum-2 instance referencing a GNSS clock on the local network. The bumps and wiggles here are clearly not random in the usual sense and rather represent the NTP clock discipline seeking a robust answer to the question: "What time is it?" That is, seeking such an answer in a situation in which neither clocks nor network can be regarded as wholly reliable. This is not the situation when a high precision GNSS reference clock is available locally, but note that NTP does an excellent job of remaining well below an offset of 1 millisecond.

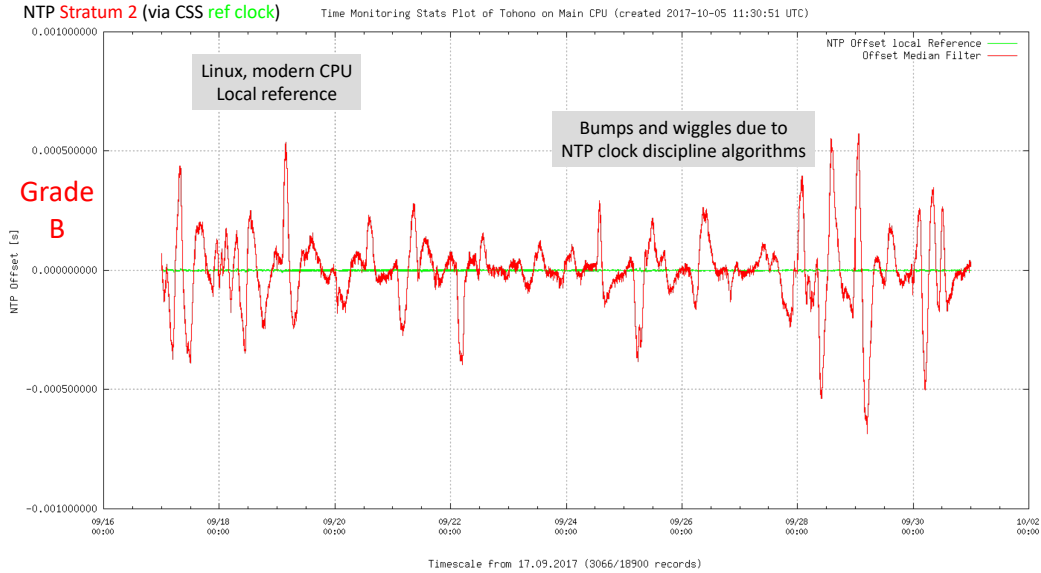

Figure 9. An NTP stratum-2 instance synchronized via a GNSS reference clock over the local network. The bumps and wiggles are indicative of the NTP clock discipline and are not random, per se. Axes are the same as Figure 7: 2 weeks and +/- 1 ms.

### 3.4 What to avoid with NTP

The Network Time Protocol has a variety of failure modes that must be guarded against. Figure 10 is also a stratum-2 NTP instance, but in this case the reference clock is not local, but rather remotely situated on the University of Arizona campus at the other end of a microwave network connection. If you look carefully you can see traces of the same NTP clock discipline hunting through this much noisier input data that now flows outside the +/- 1 millisecond window. Axes are otherwise as above.

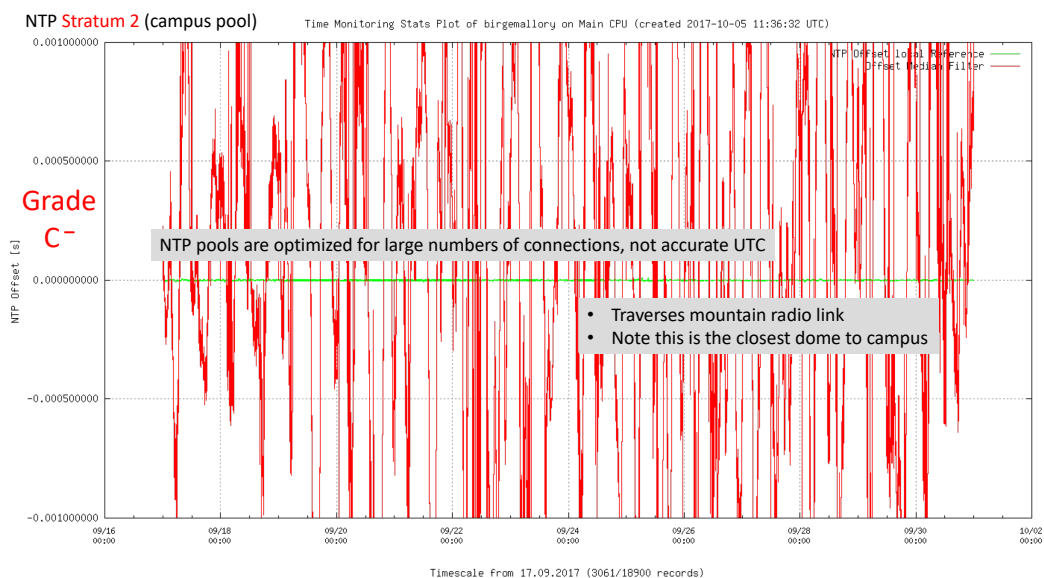

Figure 10. An NTP stratum-2 instance synchronized over a remote radio link. Axes as above.

In the absence of a local reference clock, the best that NTP can do is a rough convolution of the two previous plots. Indeed, the bumps and wiggles of the NTP clock discipline exist precisely to handle such noisy and sometimes unreliable input data. It is also worth mentioning that the usual NTP campus pool configuration, where most clocks are stratum-3 selecting from a round-robin of stratum-2 reference clocks results in client machines bouncing from one sometimes reference clock to another as loads and network weather vary. This is not a recipe for making serious scientific timekeeping measurements.

Figure 11 is back to showing a stratum-1 reference, but in this case a low-cost GPS time server. The gaps exist because this GPS device is not actually a clock, as it does not contain an oscillator and has zero holdover when the antenna or network links lose synchronization. All plots so far have been median filtered. The plot on the right of figure 11 shows the noisy raw data of such a device precisely because it lacks tempering through an oscillator's time constant.

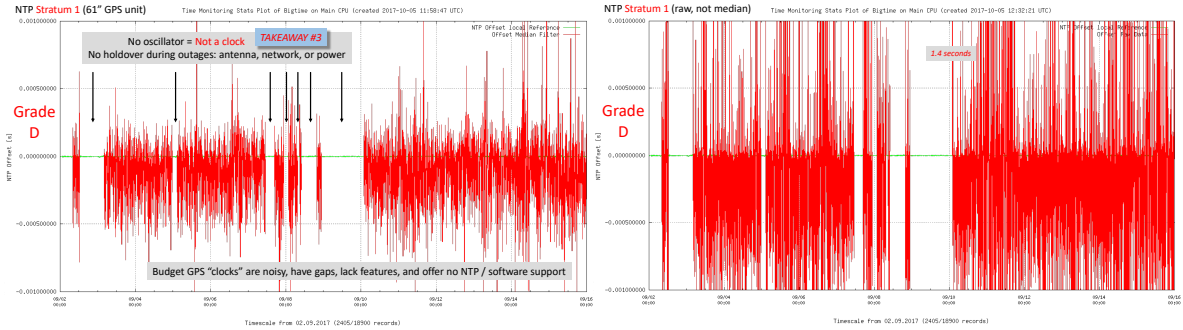

Figure 11. On the left is a local stratum-1 server relying on a low-cost GPS time server. This is not a clock, per se, since the device does not include an oscillator. When the device loses synchronization with the GPS satellites, or a variety of other network or power outage conditions it ceases to provide time. On the right is raw data from the same clock. Previous figures were all median filtered.

By contrast, a Temperature-Compensated Crystal Oscillator (TCXO) or Oven-Controlled Crystal Oscillator (OCXO) or Rubidium Atomic Clocks (for high precision applications) in good quality devices provide significant holdover time during which the devices will continue to deliver accurate time even when synchronization is lost.

Figure 12 has been zoomed out by a factor of a hundred to show the much larger excursions of a stratum-3 clock that happens to be embedded in our third-party Telescope Control System (TCS). During this one-week period there were clear diurnal swings due perhaps to loading issues as the telescope is actually slewing at night, or perhaps to interactions between the NTP clock discipline and the stringent constraints of the real-time operating system running the TCS. The gaps here were periods when the TCS was shutdown, but note that the clock was close to the green reference time with each restart but then immediately steered itself far away. Large excursions of a half-second are visible during nighttime operations, likely corresponding to telescope servo jumps.

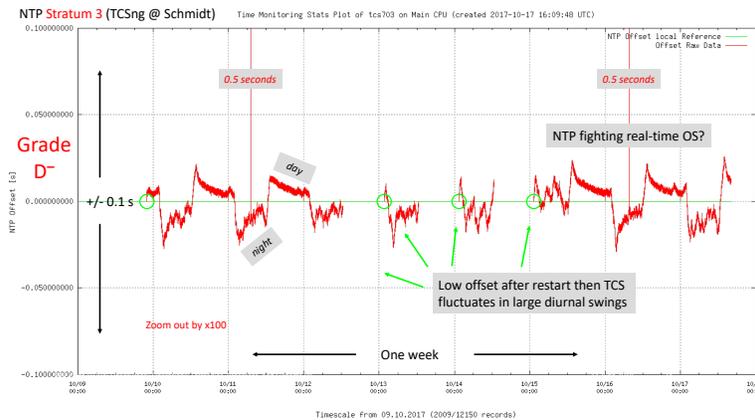

Figure 12. The stratum-3 clock within a telescope control system. Y-axis is 100x more coarse than previous plots. Distinctive diurnal pattern is perhaps due to day/night differences in loading of TCS system, e.g., the NTP clock discipline fighting against a real-time OS. Note that after a restart the device has a small offset that is driven to large excursions.

### 3.5 Use hardware time capture

It is always better to measure time pertaining to a scientific measurement than to make an estimation. Generally this means arranging for hardware time capture of pulses associated with the changing state of the instrumentation, rather than trying to constrain the time using a system clock. Figure 13 shows the stratum-0 reference corresponding to the stratum-1 NTP clock shown in figure 8. Hardware time capture of the CSS camera shutter is relative to this ruler flat trace, devoid of visible spikes over a two-week period when all the other variations demonstrated the usual jumps, bumps, and wiggles of operating telescopes on remote mountaintops. The point here is not that Catalina worked hard to achieve this result, but rather that this is the quality of result from simply investing in modern commodity timekeeping equipment.

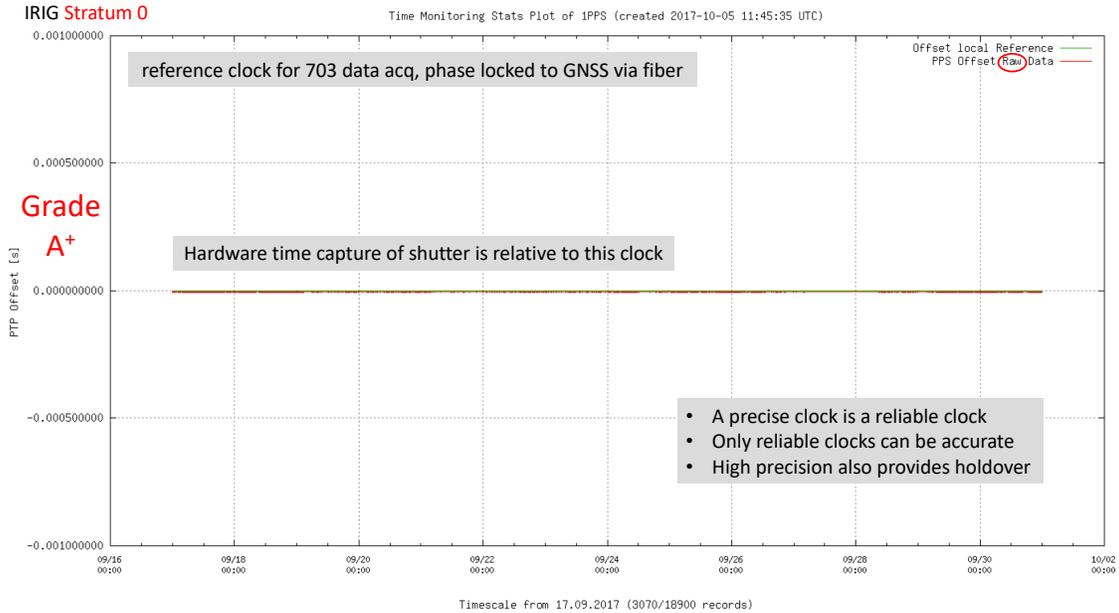

Figure 13. The stratum 0 reference clock on the IRIG-B PCIe card in one of the Catalina Sky Surveys's data acquisition computers. Hardware time capture of shutter timestamps is carried out relative to this ruler straight time signal. Note that this plot, too, displays raw data. IRIG-B cards are spec'ed for +/- 5 $\mu$s, reference clock from same vendor for +/- 50 ns.

## 4. SUMMARY

This document has discussed a number of issues associated with reliably capturing precision time measurements in remote locations such as astronomical observatories. The Catalina Sky Survey has deployed equipment as described and recommends a similar strategy relying on straightforward precision timekeeping equipment from multiple vendors to address timekeeping requirements head-on. There are three main take-aways from this paper:

### 4.1 Measure time like any other scientific quantity

Use hardware time capture to directly measure when things happen. Software interfaces generally do not guarantee accurate estimates.

### 4.2 Monitor the reliability and accuracy of your clocks

Use local NTP servers; arrange a monitoring regime as with system quantities such as diskspace, power, network, etc.

### 4.3 Not all GPS devices are actually clocks

Unless a device contains an oscillator it is not a clock. A cheap GPS "clock" may not be a clock at all.